\documentclass[12pt,reqno]{article}
\usepackage[tbtags]{amsmath}
\usepackage{amsfonts,amssymb,mathrsfs,amscd,comment}

\usepackage{epsf}\usepackage{amssymb} \usepackage{cite}
\usepackage{graphicx}
\usepackage{bm}

\newcommand{\nc}{\newcommand}

\nc{\rnc}{\renewcommand}
\nc{\beq}{\begin{equation}}
\nc{\eeq}{\end{equation}}
\nc{\eeqa}{\end{eqnarray}}

\newcommand{\la}{\langle}
\newcommand{\ra}{\rangle}

\newcommand{\ee}{\end{equation}}
\newcommand{\ba}{\begin{eqnarray}}
\newcommand{\ea}{\end{eqnarray}}

\newcommand{\R}{\mathbb R}

\begin{document}

\title{Superpositions of coherent states \\ determined by Gauss sums}

\author
{Vyacheslav~P.~Spiridonov}

\date{}

\maketitle

{Bogoliubov Laboratory of Theoretical Physics,
JINR, Dubna, Moscow Region, Russia and Laboratory for Mirror Symmetry, NRU HSE, Moscow, Russia


\begin{abstract}
We describe a family of quantum states of the Schr\"odinger cat type as superpositions of the harmonic oscillator coherent states with coefficients defined by the quadratic Gauss sums. These states emerge as eigenfunctions of the lowering operators obtained after canonical transformations of the Heisenberg-Weyl algebra associated with the ordinary and fractional Fourier transformation. The first member of this family is given by the well known Yurke-Stoler coherent state.
\end{abstract}

\maketitle

Coherent states were introduced by Schr\"odinger at the very early days of quantum mechanics \cite{Sch1}.
They formed a basis of quantum optics and helped to treat many quantum systems \cite{GK}.
However, in the discussion of famous thought possibility to have a superposition of alive
and dead cat states \cite{Sch2}, Schr\"odinger did not suggest to use superpositions of coherent
states for an experimental realization of this idea. A beautiful such quantum cat state was suggested by
Yurke and Stoler in \cite{YS1} as a result of time evolution of the standard harmonic oscillator
coherent states in a Kerr medium with nonlinear susceptibility.
It appears that this state is only the first member of an infinite family of similar superpositions
having the same group-theoretical origin. These states emerge as eigenfunctions of the lowering
(annihilation) operators after canonical transformations associated with the
ordinary and  fractional Fourier transformations. Their superposition coefficients
are determined by the quadratic Gauss sums \cite{BEW}---the remarkable number theoretic analogues
of the exactly computable Gauss integrals.

Symmetries are crucial in the formulation of fundamental
laws of physics and geometry. The beautiful objects like Platonic solids
and regular polygons are directly related to roots of unity (solutions of the algebraic equation $x^n=1$),
since they describe irreducible representations of any cyclic group. Even such a technical engineering
problem as signal processing uses roots of unity in the framework of finite Fourier transformation of
discrete signal samples. Quadratic Gauss sums are beautiful sums of the roots
of unity depending on two integer parameters and admitting exact evaluation. They have found
prominent application in the description of the Talbot effect \cite{BK,FP}---a self-imaging
phenomenon of classical optics. Another interesting use of these sums allows for a real
experimental physics approach to factorization of integers \cite{MMAMS}.  Gauss
sums emerge also in the theory of $q$-orthogonal polynomials \cite{SZ}, which
have many physical applications, but their connection to superpositions of coherent states
considered earlier in \cite{S1} was not recognized at that time.

The quantum harmonic oscillator is a fundamentally important physical model related to
the Heisenberg-Weyl algebra---a basic system with exact description of all its properties.
The first Schr\"odinger cat states, given by even and odd eigenstates of the square of annihilation
operator of the harmonic oscillator $a^2$, were constructed by Dodonov, Malkin and Man'ko in \cite{DMM}.
Although the Yurke and Stoler state  \cite{YS1} is also an eigenstate of $a^2$, it involves nontrivial
phase factors in the superposition corresponding to the rotation angle $\pi/4$. Such states are not simply
useful for a description of quantum mechanics mysteries, but they are important for solving rather
technical problems such as precision metrology \cite{Zurek} and quantum information processing \cite{GK}.

The material given in the next few paragraphs can be found in any textbook on quantum mechanics.
Still, we recall it here to make the presentation self-contained.
For simplicity, we work with the uniform normalization of physical units $\hbar =\omega=m=1$.
Then the harmonic oscillator Hamiltonian takes the form
\begin{equation}
L=\tfrac{1}{2}\left(p^2+x^2\right), \quad [x,p]=xp-px=\textup{i}, \quad \textup{i}^2=-1.
\label{ham}\end{equation}
The standard factorization of this quadratic combination of the operators of coordinate $x$ and momentum $p$ is
\begin{equation}
L=a^+a + \tfrac{1}{2},\quad a^+= \frac{-\textup{i}p+x}{\sqrt{2}}, \quad a= \frac{\textup{i}p+x}{\sqrt{2}}.
\label{fact}\end{equation}
Operators $L, a, a^+$ form the Heisenberg-Weyl algebra:
\begin{equation}
[a, a^+]=1, \qquad [L,a]=-a,\qquad [L,a^+]=a^+.
\label{HWA}\end{equation}
The ground state $|0\ra$ of the oscillator, or the vacuum, is defined as the normalized zero mode of $a$:
$a|0\ra=0, \;\la 0|0\ra=1.$ It clearly minimizes the energy of the system.
The complete system of Hamiltonian eigenstates, $|n\ra, n=0, 1, 2,\ldots$,
is easily derived by purely algebraic means:
$$
L|n\ra=\lambda_n|n\ra,\quad \lambda_n=n+\tfrac{1}{2},\quad |n\ra=\frac{1}{\sqrt{n!}}(a^+)^n\,|0\ra,
$$
with the orthonormality relation $\la n|m\ra=\delta_{nm}$ and the following action of the raising
$a^+$ and lowering $a$ operators
$$
a^+|n\ra=\sqrt{n+1}\,|n+1\ra,\qquad a|n\ra=\sqrt{n}\,|n-1\ra.
$$

In the coordinate representation, we have $x\in\mathbb{R}$ and $p=-\textup{i} d/dx$ with
$$
L=\frac{1}{2}\left(-\frac{d^2}{dx^2}+x^2\right),
\qquad L\psi_n(x)=\lambda_n\psi_n(x),
$$
where the eigenfunctions
\begin{equation}
\psi_n(x)=\la x |n\ra=\frac{H_n(x)}{\sqrt{2^nn!\sqrt{\pi}}}e^{-x^2/2}
\label{eigenfunctions}\end{equation}
involve the Chebyshev-Hermite orthogonal polynomials
$$
H_n(x)=(-1)^n e^{-x^2}\frac{d^n}{dx^n}\, e^{x^2}.
$$

Coherent states \cite{GK} are defined as eigenfunctions of the lowering operator $a$,
\begin{equation}
a|\alpha\ra=\alpha|\alpha\ra, \quad \alpha\in\mathbb{C}, \qquad \la\alpha|\alpha\ra=1.
\label{CStates}\end{equation}
Initially they were defined by Schr\"odinger \cite{Sch1} as the states for which the uncertainty relation
$\Delta x\Delta p\geq 1/2$ is minimized. However, this minimality condition does not determine
them uniquely --- the squeezed states also have such a property \cite{GK}.
An equivalent definition  uses the action of general Heisenberg-Weyl group element
on the vacuum
\begin{equation}
|\alpha\ra=e^{\alpha a^\dagger-\alpha^*a}|0\rangle=e^{-\tfrac{1}{2}|\alpha|^2}\sum_{n=1}^\infty\frac{\alpha^n}{\sqrt{n!}}|n\ra.
\label{CShwa}\end{equation}
In the coordinate representation, we have
\begin{equation}
\psi_\alpha(x)=\la x|\alpha\ra=
\frac{1}{\pi^{1/4}} e^{-\tfrac{1}{2}|\alpha|^2-\tfrac{1}{2}\alpha^2 +\sqrt{2}\alpha x -\tfrac{1}{2}x^2 }.
\label{CScoord}\end{equation}

The factorization \eqref{fact} is highly non-unique. We can write $L= A^+A+\frac{1}{2}$, where
$$
A^+= \frac{-\textup{i}p+x}{\sqrt{2}}U(p,x), \quad A=V(p,x)\frac{\textup{i}p+x}{\sqrt{2}}
$$
with the operators $U$ and $V$ satisfying the relation $UV=1$. If we
require the operator $A$ to be Hermitian conjugate of $A^+$, $A=(A^+)^\dag$,
then we have $V=U^\dag$ and $UU^\dag=1$. We assume for simplicity that the operator
$U$ is unitary, $U^\dag=U^{-1}$, and require that we have a canonical transformation  
$[A, A^+]=[a,a^+]=1$, i.e.,
\begin{equation}
U(p,x)^{-1}(p^2+x^2)U(p,x)=p^2+x^2.
\label{inv}\end{equation}
The simplest operators $U$ satisfying this equation are the parity operator,
$$
U=U^\dag=P, \quad Px=-xP,\quad Pp=-pP, \quad P^2=1,
$$
and the Fourier transformation operator \cite{M}
$$
[{\mathcal F^{\pm1}}f](y):=\frac{1}{\sqrt{2\pi}}\int_{-\infty}^\infty e^{\pm\textup{i} yx}f(x) dx.
$$
It follows that ${\mathcal F}$ is a square root of the parity transformation,
${\mathcal F}^2=P,\, {\mathcal F}^4=1,$ generating symplectic reflection
${\mathcal F} x{\mathcal F}^{-1}=p, \;{\mathcal F} p {\mathcal F}^{-1}=  -x$,
which preserves the commutation relation $xp-px=\textrm{i}$.

Since the algebra \eqref{HWA} is preserved, we can apply all
formulae \eqref{ham}-\eqref{CScoord} to the operators $A$ and $A^+$.
Consider coherent states for the transformed lowering operator $A=U^{-1}a$,
\begin{equation}
A|\alpha\ra_U =\alpha|\alpha\ra_U, \qquad  \textrm{or} \qquad a|\alpha\ra_U =\alpha U|\alpha\ra_U.
\label{CSnew}\end{equation}
As shown in \cite{S1}, for $U=P$ we obtain the following Schr\"odinger cat state
\begin{equation}
|\alpha\rangle_P=\frac{1}{\sqrt2}\left(e^{- \frac{\pi\textup{i}}{4}}|\textup{i}\alpha\rangle
+e^{\frac{\pi\textup{i}}{4}}|-\textup{i}\alpha\rangle\right).
\label{YS}\end{equation}
In the coordinate representation $\psi_\alpha^{(P)}(x)=\la x|\alpha\ra_P$,
$$
\psi_\alpha^{(P)}(x)
=\frac{\sqrt2 }{ \pi^{1/4}} \exp\left(\frac{\alpha^2-|\alpha|^2-x^2}{2}\right)
\cos(\sqrt2 \alpha x - \frac{\pi}{4}).
$$
After replacing $\alpha$ by $-\textup{i}\alpha$ we obtain the Yurke-Stoler coherent state \cite{YS1}.
Since $A^2=PaPa=-a^2$, $|\alpha\rangle_P$ is an eigenstate of the square of the
annihilation operator $a^2$ (even and odd coherent states of  \cite{DMM} have the form
$|\alpha\rangle_\pm\propto|\alpha\rangle\pm |-\alpha\rangle$).

The described group-theoretical origin of the Yurke-Stoler state was uncovered in \cite{S1}
from the $q=-1$ choice for $q$-harmonic oscillator coherent states
constructed in that paper. For general values $0<q^2<1$, the deformation of the 
harmonic oscillator potential in this
$q$-oscillator model is very complicated. An amazing fact is that the annihilation operator $a_q$ of
this model lowers the energy for $\lambda<0$, but for the $\lambda>0$ continuous spectrum states
$a_q$ is the raising operator. And vice versa, the creation operator $a_q^+$ raises the energy for $\lambda<0$
and lowers it for $\lambda>0$. As a result, both operators $a_q$ and $a_q^+$, satisfying
the relation $a_qa^+_q-q^2a_q^+a_q=1$, have normalizable eigenstates. The corresponding set
of coherent states is not investigated from the phenomenological standpoint yet.

For  $U={\mathcal F}^{\pm 1}$ the definition \eqref{CSnew} yields integro-differential equations
\begin{equation}
\left(\frac{d}{dx}+x\right)\psi_\alpha^{({\mathcal F}^{\pm 1})}(x)=\frac{\alpha}{\pi^{1/2}}\int_{-\infty}^{\infty}
e^{\pm \textup{i}xy}\psi_\alpha^{({\mathcal F}^{\pm 1})}(y)dy,
\label{FTCTdef}\end{equation}
where $x, y\in\R, \, \alpha\in \mathbb{C}$.
Hamiltonian eigenstates $|n\ra$ diagonalize Fourier transformation,
${\mathcal F}|n\ra=\textup{i}^n |n\ra.$
From this fact we can derive the following  state  \cite{S2}
\begin{equation}
|\alpha\rangle_{\mathcal F}=\tfrac{1}{2}\big( |e^{\frac{\pi \textup{i}}{4}}\alpha\ra
+e^{\frac{5\pi \textup{i}}{4}}|e^{\frac{3\pi \textup{i}}{4}}\alpha\ra
+|e^{\frac{5\pi \textup{i}}{4}}\alpha\ra
+e^{\frac{\pi \textup{i}}{4}}|e^{\frac{7\pi \textup{i}}{4}}\alpha\ra\big),
\label{FTCS}\end{equation}
which resembles a compass by the rotations angles of $\alpha$ and is similar to a state discussed in \cite{Zurek,ATP}.
Coherent states $|\alpha\rangle_{{\mathcal F}^{-1}}$  are obtained from $|\alpha\rangle_{\mathcal F}$
simply by the replacement of $ \textup{i}$ by $- \textup{i}$. In the coordinate representation
\begin{eqnarray*} &&
\psi_\alpha^{({\mathcal F})}(x)=e^{-\frac{1}{2}(|\alpha|^2+x^2)}
\big(e^{-\frac{\textup{i}}{2}\alpha^2}\cosh ((1+\textup{i})\alpha x)
+e^{\frac{\textup{i}}{2}\alpha^2+\frac{\pi\textup{i}}{4}}
\sinh ((1-\textup{i})\alpha x) \big)
\end{eqnarray*}
and $\psi_\alpha^{({\mathcal F}^{-1})}(x)=(\psi_{\alpha^*}^{({\mathcal F})}(x))^*$.
From the general mathematical standpoint, it would be interesting to construct solutions
of equations \eqref{FTCTdef} lying outside the $L^2(\mathbb{R})$ Hilbert space.

Both described choices of the operator $U$ represent particular cases of the
evolution of the harmonic oscillator in time,
$$
\textup{i}\frac{d}{dt}\psi(t)=L\psi(t),\qquad  \psi(t)=e^{-\textup{i} t L} \psi(0).
$$
For our needs we take the operator
\begin{equation}
U(\varphi)=e^{-\textup{i} \varphi(L-1/2)}
\label{U}\end{equation}
with a formal real parameter $\varphi$. Then, the transformation
\begin{eqnarray*}
U(\varphi)^{-1}x U(\varphi) = x \cos \varphi+ p \sin \varphi,
\quad
U(\varphi)^{-1} p U(\varphi) =-x \sin \varphi+ p \cos \varphi,
\end{eqnarray*}
evidently satisfies equality \eqref{inv}.  In terms of $a^+$ and $a$ operators
\begin{eqnarray*}
a(\varphi)=U(\varphi)^{-1} a U(\varphi)=e^{-\textup{i}\varphi}a,
\quad
a^+(\varphi)=U(\varphi)^{-1} a^+ U(\varphi)=e^{\textup{i}\varphi}a^+.
\end{eqnarray*}

Eigenstates of the operator $A=U(\varphi)^{-1} a$ are determined by the  integro-differential equation
\begin{eqnarray}
\left(\frac{d}{dx}+x\right)\psi_\alpha^{(\varphi)}(x)= e^{-\textup{i} \varphi(L-1/2)}\psi_\alpha^{(\varphi)}(x)
:=\sqrt{2}\alpha\int_{-\infty}^{\infty} \mathcal{K}(x,y;\varphi)\psi_\alpha^{(\varphi)}(y)dy,
\label{GenEq}\end{eqnarray}
where the kernel $\mathcal{K}$ is an analytical continuation of the Mehler kernel (the Green's function or
the propagator) and it has the form \cite{MK}
\begin{eqnarray}
\mathcal{K}(x,y;\varphi)=\frac{\mu(\varphi)} {\sqrt{2\pi |\sin \varphi}|}
\exp\big[ \textup{i}\frac{(x^2+y^2)\cos \varphi -2xy}{2\sin \varphi}\big],
\label{Green}\end{eqnarray}
with $\mu(\varphi)=e^{\textup{i}(\frac{\varphi}{2}-\frac{\pi}{4}\mathrm{sgn}(\sin\varphi))}$.
The integral transform standing on the right-hand side of equality \eqref{GenEq} is called the fractional
Fourier transformation in the literature on signal processing \cite{M}. It coincides with the
standard Fourier transform for $\varphi =-\pi/2$ and its inverse for $\varphi =\pi/2$.

Representing $|\alpha\rangle_U\equiv |\alpha\rangle_\varphi$ as a series over the Hamiltonian
eigenstates $|n\ra$, we can easily find the normalizable solution of equation \eqref{GenEq}
\begin{equation}
|\alpha\rangle_{\varphi}=e^{-\frac{1}{2}|\alpha|^2}
\sum_{n=0}^\infty \frac{\alpha^n}{\sqrt{n!}} e^{-\textup{i} \varphi \frac{n(n-1)}{2}}|n\ra.
\label{CSU}\end{equation}
For generic $\varphi$, this is a particular case of the irreducible Titulaer-Glauber coherent state \cite{TG},
which is characterized by the replacement of the coefficients $e^{-\textup{i} \varphi \frac{n(n-1)}{2}}$
by arbitrary phase factors $e^{\textup{i} \theta_n}$. However, we consider the cases
\begin{equation}
\varphi=2\pi\frac{M}{N},\quad 0 < M < N,
\label{fracphase}\end{equation}
where $M$ and $N$ are arbitrary coprime integers, i.e., $(M,N)=1$.
Then $U^N=1$ and $|\alpha\rangle_{\varphi}$ reduces to a finite superposition of $|\alpha\ra$
states, which were described above for $N=2$, when $U=P$, and $N=4$, when $U={\mathcal F}^{\pm 1}$.
For $N>2$ the states $|\alpha\rangle_{\varphi}$ are composed of more than two coherent
states, they are therefore sometimes called Schr\"odinger kittens \cite{T,ATP}.
This is different from the terminology suggested in \cite{OTLG}, where the term ``kittens''
was referring to the Schr\"odinger cat states with small coherence amplitudes.

We now describe the general family of such Schr\"odinger kitten states using
the primitive roots of unity. For $\varphi=2\pi M/N$, we have
\begin{equation}
A^N=(U^{-1}a)^N
=e^{-\textup{i}\frac{N(N-1)}{2}\varphi} a^N= \mu_N a^N,
\label{aN}\end{equation}
where $\mu_N=-1$ for even $N$ ($M$ is odd in this case) and $\mu_N=1$ for odd $N$.
As a result, we have to expand the state $|\alpha\rangle_{\varphi}$
over the eigenstates of the operator $a^N$ with the eigenvalue $\mu_N\alpha^N$:
$$
|\alpha\rangle_{\varphi}=
\sum_{k=0}^{N-1}c_k\, |e^{\frac{2\pi\textup{i}}{N}k}\alpha\rangle,\;\qquad \quad  N\; \textrm{odd},
$$
or
$$
\quad |\alpha\rangle_{\varphi}
=\sum_{k=0}^{N-1}c_k\, |e^{\frac{2\pi\textup{i}}{N}k}e^{\frac{\pi\textup{i}M}{N}}\alpha\rangle,
\; \quad N\; \textrm{even}.
$$
Using series expansions \eqref{CShwa} and \eqref{CSU} on both sides of these equalities,
we obtain
\begin{equation}
\sum_{k=0}^{N-1}c_k\, e^{\frac{2\pi\textup{i}}{N}kn}=
e^{-\pi\textup{i}\frac{M}{N} n(n-1)}, \;\quad  N\; \textrm{odd},
\label{eqs1}\end{equation}
or
\begin{equation}
\sum_{k=0}^{N-1}c_k\, e^{\frac{2\pi\textup{i}}{N}kn}
= e^{-\pi\textup{i}\frac{M}{N} n^2},  \;\quad\quad N\;  \textrm{even}.
\label{eqs}\end{equation}
In both cases we have the discrete Fourier transformation \cite{M}. Applying the inverse discrete
Fourier transform, we find the needed coefficients
\begin{equation}
c_k= \frac{1}{N}\sum_{\ell=0}^{N-1} e^{-\pi\textup{i}\frac{M\ell(\ell-1)+2k\ell}{N}},
\; \quad N\; \textrm{odd},
\label{ck1}\end{equation}
or
\begin{equation}
c_k=\frac{1}{N}\sum_{\ell=0}^{N-1} e^{-\pi\textup{i}\frac{M\ell^2+2k\ell}{N}},\; \qquad N\;  \textrm{even}.
\label{ck}\end{equation}
The emerging sums are known as the quadratic Gauss sums \cite{BEW}, or they can be called the
discrete (cyclotomic) Jacobi theta functions. Since the corresponding
coefficients  are periodic, $c_k=c_{k+N}$, all values of these sums emerge in the expansions.

It turns out that the derived finite series \eqref{ck1} and \eqref{ck} can be summed to 
closed-form expressions. First, we note that the summands are periodic with respect to the
shifts $\ell\to \ell+N$. Therefore, the sums do not change if we shift the summation index
$\ell \to \ell+m$ for an arbitrary integer $m$. Second, we define an integer $d$ as a solution
of the relation
$$
Md=1 \!\mod N, \quad d\in\mathbb{Z}.
$$
This allows completing the squares in the summand exponentials. For even $N$, this yields
$$
c_k=\frac{1}{N}\sum_{\ell=0}^{N-1} e^{-\pi\textup{i}\frac{M}{N}((\ell+kd)^2 -k^2d^2)}
=\frac{1}{N}e^{\pi\textup{i}\frac{M}{N}k^2d^2}\sum_{\ell=0}^{N-1} e^{-\pi\textup{i}\frac{M}{N}\ell^2}.
$$
Computation of the remaining standard Gauss sum is described in many places (see, e.g. \cite{BEW}).
As a result, we obtain the following general answer for an arbitrary even $N$
\begin{equation}
c_k=\frac{1}{\sqrt{N}}e^{\pi\textup{i}\frac{M}{N}\left(d^2k^2 - \frac{N}{4} \right)}\left(\frac{N}{M}\right),
\label{genEvenN}\end{equation}
where $\left(\frac{N}{M}\right)$ is the Legendre-Jacobi symbol. For arbitrary integers
$a, b>0$, $(a,b)=1$, and the prime numbers decomposition $b=p_1\cdots p_k$ we have
$\big(\frac{a}{b}\big)=\big(\frac{a}{p_1}\big)\cdots \big(\frac{a}{p_k}\big)$, where
 $\big(\frac{a}{p_i}\big)=1$, if there is an integer $x$ such that $a=x^2 \!\mod p_i$, and
$\big(\frac{a}{p_i}\big)=-1$, if there is no such $x$.

For arbitrary odd $N$ and even $M$ we have
$$
c_k=\frac{1}{N}\sum_{\ell=0}^{N-1} e^{-\pi\textup{i}\frac{M}{N}[(\ell+d(k-\frac{M}{2}))^2 -d^2(k-\frac{M}{2})^2]}
=\frac{1}{N} e^{\pi\textup{i}\frac{M}{N}d^2(k-\frac{M}{2})^2}
\sum_{\ell=0}^{N-1} e^{-\pi\textup{i}\frac{M}{N}\ell^2}.
$$
Applying again the standard Gauss sum evaluation for an odd denominator $N$ \cite{BEW}, we obtain
\begin{equation}
c_k=\frac{1}{\sqrt{N}}e^{\pi\textup{i}\left(d^2(k-\frac{M}{2})^2\frac{M}{N} +\frac{N-1}{4}\right)} \left(\frac{M}{N}\right).
\label{genOddN1}\end{equation}

Finally, when both integers $N$ and $M$ are odd, we have
$$
c_k=\frac{1}{N}\sum_{\ell=0}^{N-1} (-1)^\ell e^{-\pi\textup{i}\frac{M}{N}[(\ell+d(k+\frac{N-M}{2}))^2
-d^2(k+\frac{N-M}{2})^2]}
$$
$$ \makebox[3em]{}
=(-1)^{d(k+\frac{N-M}{2})}\frac{1}{N} e^{\pi\textup{i}\frac{M}{N}d^2(k+\frac{N-M}{2})^2}
\sum_{\ell=0}^{N-1} (-1)^\ell e^{-\pi\textup{i}\frac{M}{N}\ell^2}.
$$
Evalutation of the last plain Gauss sum with the sign alternation of terms is described in \cite{C}.
As a result, we  obtain
\begin{equation}
c_k=\frac{(-1)^{d(k+\frac{N-M}{2})}}{\sqrt{N}}e^{\pi\textup{i}\left(d^2(k+\frac{N-M}{2})^2\frac{M}{N} +\frac{N-1}{4}\right)} \left(\frac{M}{N}\right).
\label{genOddN2}\end{equation}
Substitution of the derived explicit expressions for $c_k$ into relations
\eqref{eqs1} and \eqref{eqs}  yields again the quadratic Gauss sums.
Evidently, the whole described computational procedure resembles evaluation of ordinary Gaussian integrals.

Now we can describe the Schr\"odinger kitten states for $N=3$ explicitly:
\begin{eqnarray} \label{2pi3} &&
|\alpha\rangle_{2\pi/3}
= \frac{1}{\sqrt{3}}\left( e^{-\frac{\pi\textup{i}}{6}}|\alpha\ra
+e^{-\frac{\pi\textup{i}}{6}}|e^{\frac{2\pi\textup{i}}{3}}\alpha\ra
+e^{\frac{\pi\textup{i}}{2}}|e^{\frac{4\pi\textup{i}}{3}}\alpha\ra\right),
\\ &&
|\alpha\rangle_{4\pi/3}
= \frac{1}{\sqrt{3}}\left( e^{\frac{\pi\textup{i}}{6}}|\alpha\ra
+e^{-\frac{\pi\textup{i}}{2}}|e^{\frac{2\pi\textup{i}}{3}}\alpha\ra
+e^{\frac{\pi\textup{i}}{6}}|e^{\frac{4\pi\textup{i}}{3}}\alpha\ra\right).
\label{4pi3}\end{eqnarray}
The full family of pentagonal Schr\"odinger kitten states appearing for $N=5$ has the form
\begin{eqnarray}\label{1N5} &&
|\alpha\rangle_{2\pi/5}
= \frac{1}{\sqrt{5}}\Big( e^{-\frac{\pi\textup{i}}{5}}|\alpha\ra
+e^{-\frac{\pi\textup{i}}{5}}|e^{\frac{2\pi\textup{i}}{5}}\alpha\ra
+e^{\frac{\pi\textup{i}}{5}}|e^{\frac{4\pi\textup{i}}{5}}\alpha\ra
-|e^{\frac{6\pi\textup{i}}{5}}\alpha\ra
+e^{\frac{\pi\textup{i}}{5}}|e^{\frac{8\pi\textup{i}}{5}}\alpha\ra \Big),
\\  \label{2N5}  &&
|\alpha\rangle_{4\pi/5}
= \frac{1}{\sqrt{5}}\Big( e^{-\frac{2\pi\textup{i}}{5}}|\alpha\ra
+|e^{\frac{2\pi\textup{i}}{5}}\alpha\ra
+e^{-\frac{2\pi\textup{i}}{5}}|e^{\frac{4\pi\textup{i}}{5}}\alpha\ra
+e^{\frac{2\pi\textup{i}}{5}}|e^{\frac{6\pi\textup{i}}{5}}\alpha\ra
+e^{\frac{2\pi\textup{i}}{5}}|e^{\frac{8\pi\textup{i}}{5}}\alpha\ra \Big),
\\ \label{3N5} &&
|\alpha\rangle_{6\pi/5}
= \frac{1}{\sqrt{5}}\Big( e^{\frac{2\pi\textup{i}}{5}}|\alpha\ra
+e^{-\frac{2\pi\textup{i}}{5}}|e^{\frac{2\pi\textup{i}}{5}}\alpha\ra
+e^{-\frac{2\pi\textup{i}}{5}}|e^{\frac{4\pi\textup{i}}{5}}\alpha\ra
+ e^{\frac{2\pi\textup{i}}{5}}|e^{\frac{6\pi\textup{i}}{5}}\alpha\ra
+|e^{\frac{8\pi\textup{i}}{5}}\alpha\ra \Big),
\\  \label{4N5} &&
|\alpha\rangle_{8\pi/5}
= \frac{1}{\sqrt{5}}\Big( e^{\frac{\pi\textup{i}}{5}}|\alpha\ra
+e^{-\frac{\pi\textup{i}}{5}}|e^{\frac{2\pi\textup{i}}{5}}\alpha\ra
-|e^{\frac{4\pi\textup{i}}{5}}\alpha\ra
+e^{-\frac{\pi\textup{i}}{5}}|e^{\frac{6\pi\textup{i}}{5}}\alpha\ra
+e^{\frac{\pi\textup{i}}{5}}|e^{\frac{8\pi\textup{i}}{5}}\alpha\ra \Big).
\end{eqnarray}
The hexagonal coherent states (similar to the benzen-like state discussed in \cite{RGPV})
appear for $N=6$; we skip their explicit description for brevity.

As usual, the real time evolution of all constructed coherent states has the form
$$
|\alpha,t\rangle_{\varphi}:=e^{-\textup{i}t L}|\alpha\rangle_{\varphi}
=e^{-\textup{i}\frac{t}{2}}|e^{-\textup{i}t}\alpha\rangle_{\varphi},
$$
which can be rigorously proved with the help of the Green's function \eqref{Green}.
The superpositions described above can be generated by the $\varphi$-time evolution of $|\alpha\rangle$
in a Kerr medium \cite{YS1,T} corresponding to the formal unitary operator
\begin{equation}
G(\varphi)=\exp\{\tfrac{\textup{i}}{2}\varphi(L-\tfrac{1}{2})(L-\tfrac{3}{2})\}, \quad
G(\varphi)^\dag=G(\varphi)^{-1}.
\label{Kerr}\end{equation}
We have $|\alpha\rangle_{\varphi}=G(\varphi)^{-1}|\alpha\rangle$ and
the automorphism of the Heisenberg-Weyl algebra
\begin{equation}
A(\varphi)=G(\varphi)^{-1}a G(\varphi)=U(\varphi)^{-1}a.
\label{CT}\end{equation}
Indeed, differentiating the first two members of equalities \eqref{CT} with respect to $\varphi$ yields
$$
\frac{dA(\varphi)}{d\varphi}=-\frac{\textup{i}}{2}G(\varphi)^{-1}[(L-\tfrac{1}{2})(L-\tfrac{3}{2}),a]G(\varphi)
=\textup{i}(L-\tfrac{1}{2})A(\varphi).$$
Integrating this equation yields the rightmost equality in \eqref{CT}.
However, a rigorous definition of $G(\varphi)$ as an integral operator acting in
a proper functional space similar to \eqref{GenEq} is not known. A similar drawback is present for
all choices $U=e^{\textup{i}P_k(L)}$, where $P_k(L)$ is an arbitrary polynomial
of $L$ of the degree $k>1$ with real coefficients. Repeating the above procedure for $k>1$
we can obtain superpositions of coherent states whose coefficients are determined by the
higher order Gauss sums which do not have explicit evaluations.

The general idea of building finite superpositions of canonical coherent states was suggested long ago \cite{BB,DS}.
Superpositions generated by the Kerr medium evolution operator \eqref{Kerr} at fractional times $\varphi=2\pi M/N$
were considered in \cite{GT,T}. However, corresponding states differ from those described above since the number of
components in the superpositions is twice bigger than we have. The correct counting was performed in \cite{TAC},
but only the case $M=1$ was treated in it and the explicit form of the superposition coefficients \eqref{genEvenN},
\eqref{genOddN1}, and \eqref{genOddN2} was not derived.

An interesting physical picture arises from the following observation.
Fractional numbers $M/N$ form an everywhere dense set of points on the interval [0, 1]. Since its Lebesgue measure
is zero, the probability that a random number taken from this interval is rational is equal to zero.
Therefore, in fact, none of the above superpositions can be created exactly in experiments with a Kerr medium.
Only some approximate realization is possible.
It is hard to preserve quantum coherence for states $|\alpha\rangle_{\varphi}$ to the distances corresponding
to half the revival time (half-period) $\varphi=\pi$, when the Yurke-Stoler state should be formed.
However, if we suppose that the accuracy of the position registration is $\delta\varphi\approx \pi/(2k+1)$
for some sufficiently large integer $k$, then we cannot distinguish Yurke-Stoler coherent state
from the one with $M=k$ and $N=2k+1$. And this picture persists further on: the higher accuracy
of the measurements, the more and more complicated superpositions  of states must be registered
near a fixed point.

From the mathematical standpoint, function \eqref{CSU}, which has the explicit coordinate
representation
\begin{equation}
\langle x|\alpha\rangle_{\varphi}=\frac{e^{-\frac{1}{2}(|\alpha|^2+x^2)}}{\pi^{1/4}}
\sum_{n=0}^\infty q^{n^2}\left(\frac{\alpha}{q\sqrt{2}}\right)^n  \frac{H_n(x)}{n!}, \quad
q=e^{-\frac{1}{2}\textup{i} \varphi},
\label{xCSU}\end{equation}
contains information about all
quadratic Gauss sums, similarly to the Jacobi theta functions on the natural boundary of convergence.
Therefore, it should be a rather interesting number theoretical object deserving further investigation.
For instance, the described Gauss sums obey the quadratic reciprocity law \cite{BEW,C} connecting 
the $c_k$ coefficients
for $\varphi=M/N$ with those for $\varphi=N/M$, the physical meaning of which is not yet clear.
Since this property is connected to the modular transformation for theta functions, we can conjecture that
the function $\langle x|\alpha\rangle_{\varphi}$ obeys some modular properties for an arbitrary $\varphi$.
The fractal nature of function  \eqref{xCSU} is seen also in the following observation. If we represent
the factor  $q^{n^2}$ as an integral over an exponential function of $n$, then the sum over $n$
is computable in the closed form, but this leads to the diverging integral.

In view of the wide applications that have been found
for the Schr\"odinger cat states \cite{GK,DMM,Zurek,ATP,OTLG,RGPV},
it is expected that the described superpositions of coherent states, whose interference coefficients
are given by the phase factors related to regular polygons, will find some interesting experimental implications.

\smallskip

{\bf Acknowledgments.}
The author is indebted to P.~K.~ Panigrahi for an inspiring invitation to lecture at the QIQT-2021 School
(Kolkata, July 2021) and to A.~S.~ Zhedanov for a useful discussion on the Gauss sums.

This study has been partially funded within the framework of the HSE University Basic Research Program.


%

\end{document}